\documentclass[aip,pop,preprint,amsmath,amssymb,nofootinbib]{revtex4-1}

\usepackage{graphicx}

\def\grl{{\itshape Geophys. Res. Lett.} }
\def\jgr{{\itshape J. Geophys. Res.} }
\def\prl{{\itshape Phys. Rev. Lett.} }
\def\pop{{\itshape Phys. Plasmas} }
\def\ssr{{\itshape Space Sci. Rev.} }

\begin{document}

\title{Electron dynamics surrounding the X-line in asymmetric magnetic reconnection}

\author{Seiji Zenitani}
\affiliation{Research Institute for Sustainable Humanosphere, Kyoto University, Gokasho, Uji, Kyoto 611-0011, Japan; Email: zenitani@rish.kyoto-u.ac.jp.}
\affiliation{National Astronomical Observatory of Japan, 2-21-1 Osawa, Mitaka, Tokyo 181-8588, Japan}
\author{Hiroshi Hasegawa}
\affiliation{Institute of Space and Aeronautical Science, Japan Aerospace Exploration Agency, 3-1-1 Yoshinodai, Chuo, Sagamihara, Kanagawa 252-5210, Japan}
\author{Tsugunobu Nagai}
\affiliation{Tokyo Institute of Technology, Tokyo 152-8551, Japan}

\date{Submitted to JGR Space Physics, MMS special issue}

\begin{abstract}
Electron dynamics surrounding the X-line in magnetopause-type asymmetric reconnection
is investigated using a two-dimensional particle-in-cell simulation.
We study electron properties of three characteristic regions
in the vicinity of the X-line.
The fluid properties, velocity distribution functions (VDFs),
and orbits are studied and cross-compared.
On the magnetospheric side of the X-line,
the normal electric field enhances
the electron meandering motion from the magnetosheath side.
The motion leads to a crescent-shaped component in the electron VDF,
in agreement with recent studies. 
On the magnetosheath side of the X-line,
the magnetic field line is so stretched in the third dimension that
its curvature radius is comparable with typical electron Larmor radius.
The electron motion becomes nonadiabatic, and therefore
the electron idealness is no longer expected to hold. 
Around the middle of the outflow regions,
the electron nonidealness is coincident with
the region of the nonadiabatic motion.
Finally, we introduce a finite-time mixing fraction (FTMF) to
evaluate electron mixing.
The FTMF marks the magnetospheric side of the X-line,
where the nonideal energy dissipation occurs.
\end{abstract}

\maketitle

\section{Introduction}

Magnetic reconnection is
a fundamental process at plasma boundary layers.
By changing the field-line topology,
the reconnection process allows
the rapid release of the stored magnetic energy to plasma energies, 
as well as
the transport of mass, momentum, and energies across the boundaries.
The physics of magnetic reconnection has long been studied
by means of magnetohydrodynamic and kinetic theories, computer simulations, and
satellite observations in solar-terrestrial environments (e.g., \citet{birn07,treumann13}).

At the Earth's dayside magnetopause,
magnetic reconnection occurs between the magnetosheath and the magnetoshere.
Plasma properties of the two regions are quite different.
The magnetosheath side is dominated by a shocked solar-wind plasma,
while the magnetospheric side is dominated by the Earth's dipole magnetic field.
To discuss magnetopause reconnection,
it is important to understand
the influence of {\it asymmetric} upstream conditions to the reconnection process.
Many aspects of asymmetric reconnection have been explored over a decade.
For discussions of earlier progress in the study of asymmetric reconnection,
readers may wish to consult
\citet{mozer11} for theoretical aspects,
\citet{paschmann13} for observational aspects, and
\citet{eastwood13} and \citet{cassak16} for both aspects.

In order to observe kinetic signatures in near-Earth reconnection sites,
NASA has embarked the Magnetospheric Multiscale (MMS) mission.\citep{burch16b}
Thanks to multi-spacecraft operation and high-resolution instruments,
MMS is able to measure electron-physics signatures
for the first time in magnetospheric physics. 
The MMS has extensively observed
current layers at the Earth's magnetopause
during the first phase of the mission in 2015--2017.
The mission has strongly motivated
theorists to explore the electron physics in asymmetric reconnection. 
In the following paragraphs,
we briefly review 
recent progress in the electron kinetic physics and the relevant signatures
in asymmetric reconnection,
by means of particle-in-cell (PIC) simulations. 

\citet{prit09a} was one of the first to focus on
the electron physics near the X-line.
They studied various quantities around the reconnection site
in their PIC simulations,
in order to identify the electron-physics region. 
Their results were successively reviewed by \citet{mozer11}. 
They only found that
asymmetric reconnection appears to be quite different from
symmetric reconnection in various parameters. 
This inconvenient fact led researchers to
seek for new parameters to interpret the results. 
\citet{zeni11c} developed
a frame-independent formula of the nonideal energy dissipation,
which identifies an electron-scale dissipation region surrounding the X-line
in various cases, including asymmetric reconnection
with and without the guide field. 
This visualized that
asymmetric reconnection involves an electron-physics layer surrounding X-line. 
Meanwhile,
\citet{egedal11} discussed the influence of the parallel electric field
to electron orbits and velocity distribution functions (VDFs).
Since the field-aligned component of the reconnection electric field
traps the electrons in the inflow regions,
the composition of the electron VDFs can be organized by using a parallel potential.
When the asymmetry in the inflow density is high,
the parallel field tends to accelerate electrons
toward the high-density magnetosheath side in an exhaust region.

\citet{hesse14} investigated
the electron physics near the X-line in detail.
They evaluated the electron Ohm's law across the reconnection layer. 
The authors found that,
unlike in symmetric reconnection,
the bulk inertial effect sustains
the reconnection electric field at the X-line
and that
the divergence of the electron pressure tensor supports
the reconnection electric field at the flow stagnation point. 
At the stagnation point, they found that
an electron VDF consists of a gyrotropic core component and
a crescent-shaped meandering component originating from the magnetosheath. 
The crescent-shaped VDF has drawn immediate attention.
The MMS spacecraft successfully measured crescent-shaped electron VDFs during
a reconnection event at the dayside magnetopause.\citep{burch16c}
Since the crescent appeared in a perpendicular plane to the magnetic field in the VDF,
they called it a ``perpendicular crescent.''
They discovered a new crescent-shaped VDF in the same event.
Since it appeared in a parallel plane,
it was named a ``parallel crescent.''
The parallel-crescent electrons are streaming away from the X-line.
The MMS observation stimulated
further investigation on the electron VDFs.

\citet{bessho16} examined
the trajectories of meandering electrons and
the detailed structure of the electron VDFs near the X-line.
Recognizing a strong normal electric field ($E_z$),
they developed a simple one-dimensional model across the reconnection layer.
Then, they discussed key signatures of the electron VDFs and motions,
such as
requirements for the crescent-shaped component and
the penetration distance of sheath-origin electrons. 
Their model successfully explained the spatial variation of the electron VDFs.
\citet{chen16} constructed
an array of electron VDFs around the reconnection site.
They discussed the relevance between these VDFs and the electron motion.
Signatures of the meandering motion of the sheath-origin electrons,
such as the acceleration in the current-carrying direction and
the slow rotation by the normal magnetic field,
were evident in the VDFs.
They reported that
the combination of
the core component from the magnetosphere and
the crescent-shaped meandering component from the magnetosheath
is a robust feature. 

\citet{shay16} examined
various plasma properties and different measures around the X-line.
They found that the normal electric field along the separatrix
touches the field reversal only near the X-line.
This led them to propose the normal electric field at the field reversal
as a handy signature of the dissipation region.
They discussed the crescent-shaped VDFs
using a one-dimensional model in a sophisticated manner. 
They pointed out that
crescent-shaped VDFs are found in many places along the separatrix. 
\citet{egedal16} studied
electron VDFs along the sphere-side separatrix in their PIC simulation. 
They showed that
gyrotropic motions of high-density, sheath-origin electrons
account for both the perpendicular and parallel crescents.
The authors argued that
the sheath-origin electrons have higher energies,
because they are accelerated by the normal electric field.
Then they are trapped along the separatrix
by the mirror force and the parallel electric field.
The trapping model predicts loss cones in the VDF.
As a result, their VDF exhibits a parallel-crescent,
a high-energy shell with a loss cone in the incoming-side.
They showed that electron diamagnetic effect also
results in a perpendicular crescent VDF. 
\citet{lapenta17} analyzed the single-particle motion
in the immediate vicinity of the field reversal.
They claimed that the normal electric field is not essential for the crescent VDFs.
They also showed crescent-shaped VDFs in symmetric reconnection as well.

Investigations with 3D PIC simulations are in progress.
\citet{prit11} reported ripple-like fluctuations
along the magnetosphere-side separatrices,
where there is a steep gradient in the plasma density.
Subsequent simulations using high mass ratios\citep{roy12,prit12,prit13}
revealed that the fluctuations are driven by
the electrostatic lower-hybrid drift instability (LHDI).
This LHDI is usually localized away from the X-line
and occurs on an electron scale:
The wave number $k_y$ in the third direction satisfies $k_y\rho_e \lesssim 1$,
where $\rho_e$ is the electron Larmor radius. 
The LHDI-driven turbulence modifies
the local momentum balance of the electron fluid,\citep{roy12,prit13,price16}
however, all these studies agree that
the turbulence does not significantly alter
the gross properties of 2D reconnection.
The crescent-shaped VDFs were reported in these 3D cases.\citep{price16,lea17}

Another issue is the electron mixing during magnetic reconnection.
In a different context, \citet{dau14} studied
the electron mixing in 3D asymmetric reconnection.
In their study, many electrons
leak out from the exhaust region to the inflow region,
due to the complex field-line geometry in 3D.
Then the authors utilized the electron mixing fronts to
discuss the reconnection rate in turbulent reconnection.
\citet{lea17} also examined
electron mixing in 3D asymmetric reconnection in antiparallel configuration.
They found that the LHDI-driven turbulence invokes
the electron mixing and parallel heating
in a magnetospheric inflow region.

Indeed, there has been significant progress in
understanding electron kinetic properties of asymmetric reconnection.
Now many scientists agree that
asymmetric reconnection involves
the crescent-shaped electron VDFs
on the magnetospheric side of the X-line,
due to the meandering motion of the sheath-origin electrons.\citep{hesse14,bessho16,chen16,shay16}
It also turned out that 
the (perpendicular) crescent is not a unique signature near the X-line,
because it is essentially a finite Larmor radius (FLR) effect
at boundary layers such as the separatrix regions.\citep{chen16,shay16,norgren16,egedal16}
The mechanism for the parallel crescent has just been proposed.\citep{egedal16}
In 3D, a growing consensus is that
a branch of LHDI invokes turbulence
on the magnetospheric side of the reconnection layer.\citep{prit11,prit12,prit13,roy12,price16,lea17}
However, despite the 3D turbulence,
many 2D properties of asymmetric reconnection remain unchanged.
The electron crescent-shaped VDF persists in 3D,
as confirmed by MMS observations.\citep{burch16c}
It is possible that
the LHDI-driven turbulence does not
destroy FLR features such as the crescent-shaped VDFs,
because the LHDI has a longer wavelength than the electron Larmor radius.
An emerging issue is the enhanced electron mixing in 3D.\citep{dau14,lea17}
However, researches on the electron mixing have just started recently.
Previous studies relied on a very simple diagnosis.
To go further, it is desirable to further develop diagnosis frameworks
to evaluate the electron mixing.

In this study, we explore detailed properties of asymmetric magnetic reconnection
by using 2D PIC simulations.
While previous researches extensively studied electron VDFs,
we examine various electron-physics signatures surrounding the X-line
from various angles, including single-particle dynamics.
The electron nonidealness and the composition of the electron VDFs
will be analyzed, with help from a dataset of many electron orbits.
We further propose a new diagnosis to evaluate the electron mixing.
This will be important for future study on turbulent 3D reconnections. 

The paper is organized as follows.
Section \ref{sec:kappa} briefly introduces
curvature parameters that indicate nonadiabatic particle motions.
Section \ref{sec:setup} describes the numerical setup of a 2D PIC simulation.
Section \ref{sec:results} presents the simulation results.
We will show electron fluid properties, electron VDFs, electron orbits, and
the electron mixing.
Section \ref{sec:discussion} contains discussions,
followed by conclusions in Section \ref{sec:conclusion}.

\section{Curvature parameters}
\label{sec:kappa}

The particle motion around a field reversal,
approximated by $\boldsymbol{B}(z)=B_0(z/L)\boldsymbol{e_{x}} + B_n \boldsymbol{e_{z}}$
on the first-order,
is characterized by a curvature parameter,\citep{BZ89}
\begin{equation}
\label{eq:BZ89}
\kappa
\equiv
\sqrt{\frac{R_{\rm c,min}}{\rho_{\rm max}}}
= \Big| \frac{B_n}{B_0} \Big| \sqrt{\frac{L}{\rho_0}}
= \sqrt{ \frac{B_n L}{B_0 \rho_n} }
\end{equation}
where $R_{\rm c}$ is the curvature radius of magnetic field lines,
$L$ is the typical length,
$\rho$ is the Larmor radius,
$\rho_0$ is the Larmor radius about the reference magnetic field $B_0$, and
$\rho_{n}$ about the normal magnetic field $B_n$.
When $\kappa \lesssim 2.5$,
magnetic moments are no longer conserved and therefore
the particle motion is considered to be nonadiabatic.
Particle motion becomes highly chaotic for $\kappa \sim 1$.
For $\kappa \ll 1$, several classes of nongyrotropic motions appear,
such as the Speiser (transient) motion and the regular motion.\citep{speiser65,chen86}

When the system has a shear field $B_s$ in the third direction, i.e.,
$\boldsymbol{B}(z) = B_0(z/L)\boldsymbol{e_{x}} + B_s \boldsymbol{e_{y}} + B_n \boldsymbol{e_{z}}$,
particle motion can be discussed by modified curvature parameters,\citep{kari90,BZ91}
\begin{equation}
\label{eq:BZ91}
\kappa_{s} \equiv \frac{B_{s}}{B_0} \sqrt{\frac{L}{\rho_{0}}}
,~
\kappa_{n} \equiv \frac{B_{n}}{B_0} \sqrt{\frac{L}{\rho_{0}}}
,~
\kappa_{\rm tot} \equiv |\kappa_n| \Big( 1+\Big|\frac{\kappa_s}{\kappa_n}\Big|^2 \Big)^{3/4}
.
\end{equation}
Here, $\kappa_s$, $\kappa_n$ and $\kappa_{\rm tot}$ are
shear, normal, and total curvature parameters.
When $B_s=0$, the last two are reduced to the classical curvature parameter,
$\kappa_{\rm tot}=|\kappa_n|=\kappa$.
Similarly, $\kappa_{\rm tot} \sim 1$ corresponds to chaotic particle motion. 
One article mentioned $\kappa_{\rm tot} = 3$ as a boundary between
adiabatic and nonadiabatic particle motion.\citep{BZ91a}
Note that particle motion is discussed in a moving frame,
in which the electric field is transformed away.

Recently, \citet{lea13} proposed
the following curvature parameter for electrons,
\begin{equation}
\label{eq:K}
{\mathcal{K}^{2}}
\equiv {\frac{R_{\rm c}}{\rho_{\rm eff}}}
=
\Biggl(
\frac{|\boldsymbol{b}\cdot\nabla\boldsymbol{b}|}{\Omega_{\rm ce}}
\sqrt{
\frac{{\rm tr}({\mathbb{P}_e})}{3m_en_e}
}
~
\Biggr)^{-1}
\end{equation}
where $R_{\rm c} \equiv |\boldsymbol{b}\cdot\nabla\boldsymbol{b}|^{-1}$ is
the magnetic curvature radius,
$\rho_{\rm eff}$ is the effective Larmor radius,
$\boldsymbol{b} \equiv \boldsymbol{B}/|B|$ is the unit vector, and
$\mathbb{P}_e$ is the electron pressure tensor.
In this paper, we call it an ensemble curvature parameter.
Note that
$\rho_{\rm eff}$ is evaluated
in the rest frame of the electron bulk flow,
while the original curvature parameters are considered
in the moving frame at the {\bf E}~$\times$~{\bf B} velocity,
$\boldsymbol{V}_{{\bf E}\times{\bf B}}$.
Therefore, when the electron thermal speed ($v_{\rm e,th}$) is large enough,
$v_{\rm e,th} \gg |\boldsymbol{V}_e - \boldsymbol{V}_{{\bf E}\times{\bf B}}|$,
the $\mathcal{K}$ parameter represents
a curvature parameter for a typical electron in the distribution.

\section{Simulation model}
\label{sec:setup}

We employed a partially implicit PIC code.\citep{hesse99}
Lengths, time, and velocities are normalized by
the ion inertial length $d_i=c/\omega_{pi}$,
the ion cyclotron time $\Omega_{ci}^{-1}=m_i/(eB_0)$, and
the ion Alfv\'{e}n speed $c_{Ai}=B_0/(\mu_0 m_i n_0)^{1/2}$, respectively.
The ion plasma frequency $\omega_{pi}=( e^2n_0/\varepsilon_0 m_i)^{1/2}$ is
evaluated for the reference density $n_0$.
We used the following initial model,
\begin{eqnarray}
\boldsymbol{B}(z) &=&
B_0 [ R + \tanh (z/L) ]~\boldsymbol{e}_x
,
\\
n(z) &=& n_0 \big[ 1 - 2 \alpha R \tanh(z/L) - \alpha \tanh^{2}(z/L) \big]
,
\end{eqnarray}
where $L=0.5 d_i$ is the half thickness of the transition layer. 
This model was proposed by \citet{prit08} and slightly modified by \citet{klimas14}. 
The $R=1/2$ parameter gives a variation in magnetic field from
$-0.5B_0$ to $1.5B_0$.
The $\alpha=1/3$ parameter
gives a density variation from $n_0$ to $(1/3)n_0$.
The plasma temperature is uniform.
The initial guide magnetic field is set to zero: $B_y=0$.
The corresponding plasma $\beta$ varies from $12$ to $4/9$ across the layer.
The mass ratio,
the ratio of the electron plasma frequency to the electron cyclotron frequency,
and the ratio of plasma temperatures are
$m_i/m_e=25$, $\omega_{pe}/\Omega_{ce}=4$, and $T_e/T_i = 0.2$, respectively. 
The simulation domain size is $x,z \in [0,128.0]\times[-12.8, 12.8]$.
It is resolved by $2000 \times 800$ grid cells.
Periodic ($x$) and reflecting ($z$) boundaries are employed.
To trigger the reconnection process,
we impose a small flux perturbation of $\delta B \approx 0.1$ near the center.
These conditions are almost the same as
used in our previous studies,\citep{zeni11c,hesse13}
except for the system size in $x$.
The domain is twice larger in $x$
to eliminate the electron circulation effects across the periodic boundaries,
as will be discussed in Section \ref{sec:mix}.
We use $1.1 \times 10^9$ particles.
During $30<t<40$,
we record self-consistent trajectories of $1.8 \times 10^7$ electrons
(three percent of the total electrons).
Since they are selected only by the particle id in the simulation,
there is no selection bias.
We analyze the ``trajectory dataset'' as well as snapshot data.

Note that our coordinate system differs from the GSM system at the magnetopause
in which $x$ and $z$ are interchanged. 
We call the high-density side ($z \ll 0$),
the magnetosheath, the sheath, or the lower half.
We also call the low-density side ($z \gg 0$),
the magnetosphere, the sphere, or the upper half.

\begin{figure*}[htpb]
\centering
\includegraphics[width={\textwidth},clip,bb=0 0 998 875]{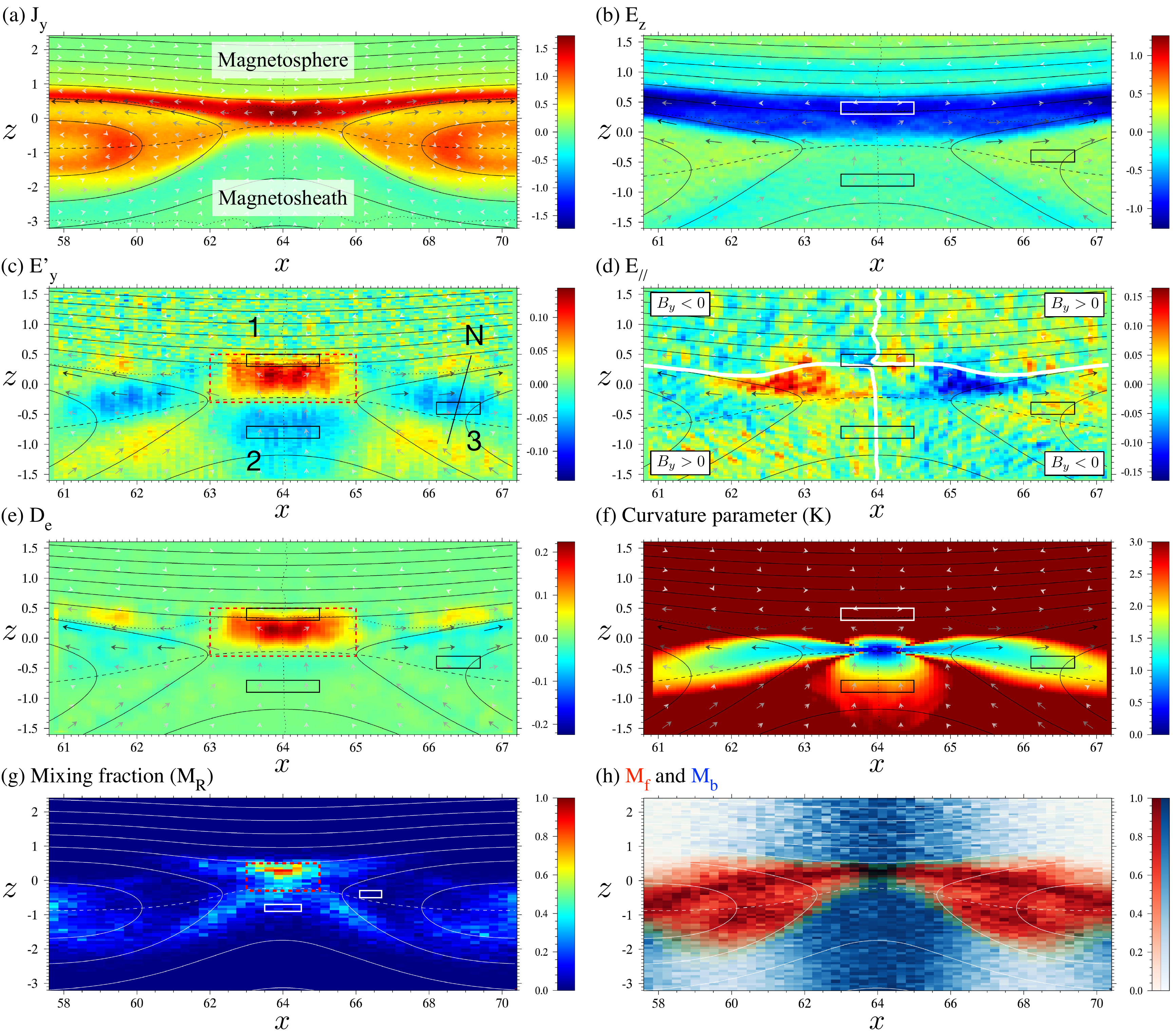}
\caption{(Color online)
\label{fig:snapshot}
Snapshots at $t=35$.
(a) The out-of-plane electric current $J_{y}$,
(b) the normal electric field $E_z$,
(c) the nonideal electric field $(\boldsymbol{E}+\boldsymbol{V}_e\times\boldsymbol{B})_y$,
(d) the parallel electric field $E_{\parallel}$ with a white contour of $B_y=0$,
(e) the energy dissipation $\mathcal{D}_e$,
(f) the ensemble curvature parameter $\mathcal{K}$ (Eq.~\ref{eq:K}),
(g) the finite-time mixing fraction $\mathcal{M}_R$ (Eq.~\ref{eq:M}), and
(h) the forward-time FTMF $\mathcal{M}_f$ in red (Eq.~\ref{eq:M_in}) and the backward-time FTMF $\mathcal{M}_b$ in blue (Eq.~\ref{eq:M_out}).
}
\end{figure*}

\section{Results}
\label{sec:results}

\subsection{Fluid properties}

Shown in Figure \ref{fig:snapshot} are selected properties at $t=35$.
They are averaged over $34.5<t<35.5$ to remove noise.
The reconnection rate peaks around this time.
The evolution of the rate in a similar configuration was presented
in \citet{hesse13} (see Fig.~4 in the paper).
In Figure \ref{fig:snapshot},
the contour lines, the dashed line, and the arrows indicate
the in-plane magnetic field lines,
the field reversal ($B_x=0$), and
the in-plane electron velocity.
Figure \ref{fig:snapshot}(a) shows
the out-of-plane electric current ($J_y$).
The electric current is intense near the X-line
and on the magnetospheric side of the reconnection layer. 

Figures \ref{fig:snapshot}(b)-(f) focus on the reconnection site.
The X-line is located at $(x,z)=(64.0,-0.2)$.
Shown in Figure \ref{fig:snapshot}(b) is the normal electric field $E_z$.
There is a strong polarization electric field $E_z<0$
along the sphere-side separatrix.\citep{prit08,tanaka08}
Its amplitude is an order-of-magnitude larger than the other components.
As pointed out by \citet{shay16},
the $E_z<0$ layer touches the field reversal
only near the dissipation region.

Figure \ref{fig:snapshot}(c) displays
the out-of-plane component of the nonideal electric field ($E'_y$),
where $\boldsymbol{E}' = \boldsymbol{E} + \boldsymbol{V}_e \times \boldsymbol{B}$.
This is related to the nonideal transport of the in-plane magnetic flux.
Three characteristic regions are indicated by the numbers.
We call them Regions 1, 2, and 3, respectively.
Region 1 is the positive-$E'_y$ region in red near the X-line.
At the X-line, $E'_y$ is weakly positive and
it becomes strong in the upper area of the field reversal. 
The second is the negative-$E'_y$ region
below the X-line ($z \lesssim -0.4$).
This region is interesting, because
there is no negative-$E'_y$ region in the magnetospheric side.
The third is the negative-$E'_y$ region
in the right outflow region ($65.5<x<67,-0.5<z<0$).
There is another one in the left outflow region ($61<x<62.5,-0.5<z<0$), but
we focus on the right one due to the left-right symmetry.
Later in this paper,
electron particle properties will be studied in the three boxes.
We call them Boxes 1, 2, and 3. 
Figure \ref{fig:snapshot}(e) shows the nonideal energy transfer,
$\mathcal{D}_e \approx \boldsymbol{J}\cdot \boldsymbol{E}'$.\citep{zeni11c}
Even though there are weak $\mathcal{D}_e > 0$ regions
on the magnetosphere-side separatrices
(the yellow region at $x=61.8$, $66.3$ and $z=0.4$),
there is an enhanced $\mathcal{D}_e > 0$ region around Region 1,
including the X-line.
We call this region the dissipation region.
The energy dissipation is basically approximated
by $\mathcal{D}_e \approx J_y E'_y$,
and therefore
the dissipation region appears to be
a product of Figures \ref{fig:snapshot}(a) and (c).
Around the X-line,
the energy dissipation is further enhanced by the $z$ components,
$J_z E'_z \approx J_z E_z > 0$.

Figure \ref{fig:snapshot}(d) shows the parallel electric field $E_{\parallel}$.
Before examining the electric field,
we briefly explain the structure of the Hall magnetic field $B_y$.
The white lines indicate $B_y=0$.
Similar to symmetric reconnection,
in-plane electron flows generate a quadrupolar $B_y$,\citep{sonnerup79}
as indicated by the small labels. 
In asymmetric reconnection, many electrons travel upward
from the magnetosheath to the magnetosphere. 
As a result, the two $B_y=0$ lines cross on the magnetospheric side from the X-line,
which is almost identical to the electron stagnation point. 
Concerning $E_{\parallel}$,
there are a localized bipolar structure around the X-line and
faint quadrupolar structures in the exhaust regions and in the inflow regions. 
In the inflow regions,
the background quadrupolar structure consists of
the $E_{\parallel}>0$ (yellow) region in the upper right,
$E_{\parallel}<0$ (light blue) in the upper left,
$E_{\parallel}>0$ in the lower left, and
$E_{\parallel}<0$ in the lower right. 
They are projections of
the reconnection electric field $E_y$ to
the Hall magnetic field $B_y$,
similarly to those in symmetric reconnection.\citep{prit01b}
Inside the exhaust regions ($z \gtrsim -0.5$), 
one can see $E_{\parallel}>0$ in the right
and $E_{\parallel}<0$ in the left
below the horizontal $B_y=0$ line.
This is because
the parallel electric fields are directed toward the high-density magnetosheath side,
to adjust the electron diffusion along the field lines.\citep{egedal11}
The bipolar $E_{\parallel}$ structure near the X-line,
the left $E_{\parallel}>0$ (red) region and the right $E_{\parallel}<0$ (blue) region,
is attributed to
the $y$-projection of the reconnection electric field $E_y$ and
the $z$-projection of the polarization field $E_z$ (Fig.~\ref{fig:snapshot}(b)). 
It is considered as
an asymmetric variant of the inner quadrupolar structure of $E_{\parallel}$
in symmetric reconnection.\citep{prit01b}

Figure \ref{fig:snapshot}(f) displays
the ensemble curvature parameter $\mathcal{K}$ for electrons (Eq.~\ref{eq:K}).
The magnetic curvature radius is computed by \citet{shen03}'s method.
In the magnetospheric side,
since the field lines are almost straight,
the parameter is large $\mathcal{K} \gg 3$. 
In contrast, in the magnetosheath side,
one can see a small-$\mathcal{K}$ region around Region 2 and
narrow $\mathcal{K}\lesssim 2$ layers in the outflow regions.
The outflow layer is slightly above the field reversal $B_x=0$. 
Figures \ref{fig:snapshot}(g) and \ref{fig:snapshot}(h) will be described
later in Section \ref{sec:mix}.

\begin{figure*}[hthp]
\centering
\includegraphics[width={\textwidth},clip,bb=0 0 520 564]{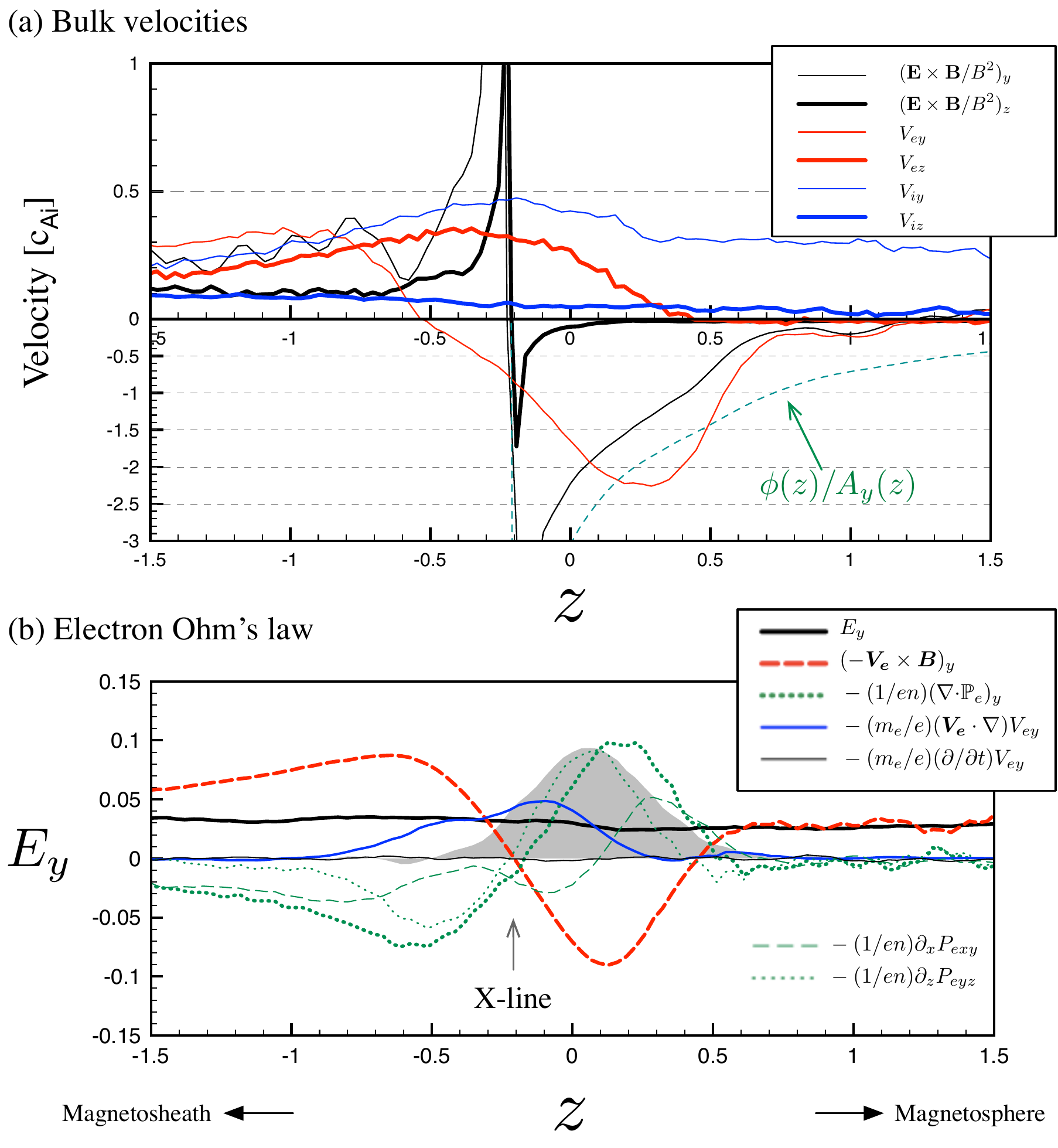}
\caption{(Color online)
\label{fig:eohm}
(a)
The 1D profile of the plasma velocities along the inflow line, $x=64.0$.
Different scales are used for the upper ($V>0$) and lower ($V<0$) halves.
The green dashed curve estimates
the influence of the normal electric field $E_z$
(the $\phi(z)/A_y(z)$ term in Eq.~\ref{eq:bessho}).
This will be discussed in Section \ref{sec:VDF}. 
(b)
The composition of the electron Ohm's law (Eq.~\ref{eq:eohm}) along the inflow line ($x=64.0$).
The pressure tensor term ($\nabla\cdot{\mathbb{P}_{\rm e}}$) is
further decomposed into the two components, shown in the bottom right.
The nonideal energy dissipation $\mathcal{D}_e$ is rescaled and overplotted in the gray shadow.
}
\end{figure*}

To understand the $z$-structure of the reconnection layer,
we show plasma velocities along the inflow line ($x=64.0$)
in Figure \ref{fig:eohm}(a).
For convenience,
we use different scales in the upper ($V>0$) and lower ($V<0$) halves.
The {\bf E}~$\times$~{\bf B} velocities become singular at the X-line ($z = -0.2$),
because the electric field remains finite. 
The ions travel in $+y$ and the electrons in $-y$
so that they carry the $y$-current (Fig.~\ref{fig:snapshot}(a)).
Aside from noises in $V_{{\bf E}\times{\bf B},y}$,
the electron bulk speed $V_{ey}$ (the thin red line) is larger than
the ion speed $V_{iy}$ (the thin blue line) and the {\bf E}~$\times$~{\bf B} speed,
$V_{ey} > V_{iy} \approx V_{{\bf E}\times{\bf B},y}$, in the magnetosheath ($z \lesssim -0.8$).
This is puzzling for the following two reasons:
First, we expect that electrons are magnetized
$\boldsymbol{V}_{e} \approx \boldsymbol{V}_{{\bf E}\times{\bf B}}$,
even when the ions are not magnetized
$\boldsymbol{V}_{i} \ne \boldsymbol{V}_{{\bf E}\times{\bf B}}$.
Second, the ions and electrons carry
the electric current in the opposite direction, $J_y<0$.
Although hardly visible in Figure \ref{fig:snapshot}(a),
there are current layers in light blue ($J_y<0$)
around Region 2 and along the magnetosheath-side separatrices.
In the $z$ direction, both ions and electrons travel upward.
The electrons travel faster than the ions in the $z \lesssim 0.3$ region,
$V_{ez} > V_{iz}$. 
At $z \lesssim -0.7$,
one can see the $z$-projection of the puzzling feature,
$V_{ez} > V_{iz} \approx V_{{\bf E}\times{\bf B},z}$. 
This $\boldsymbol{V}_{e} \ne \boldsymbol{V}_{{\bf E}\times{\bf B}}$ region
corresponds to Region 2,
as will be discussed in Section \ref{sec:discussion}.

Figure \ref{fig:eohm}(b) displays
the composition of the reconnection electric field $E_y$
along the inflow line ($x=64.0$).
The gray shadow indicates a rescaled value of
the nonideal energy dissipation $\mathcal{D}_e$ (Fig.~\ref{fig:snapshot}(e)).
The electric field $E_y$ is decomposed by using the electron Ohm's law,
\begin{equation}
\boldsymbol{E} 
=
-
{\boldsymbol{V}_e}\times\boldsymbol{B}
-
\frac{1}{en} {\nabla \cdot} {\mathbb{P}_e}
-
\frac{m_e}{e} \Big( (\boldsymbol{V}_e\cdot\nabla)\boldsymbol{V}_e+\frac{\partial \boldsymbol{V}_e}{\partial t}\Big)
.\label{eq:eohm}
\end{equation}
The electron pressure tensor term ($\nabla\cdot\mathbb{P}_e$),
presented in green, is further decomposed into
the $\partial_x P_{exy}$ term (the dashed line)
and the $\partial_z P_{eyz}$ term (the dotted line).
In symmetric reconnection,
the reconnection electric field is balanced by
the pressure tensor term at the X-line,\citep{hesse99}
however, other terms can balance it in asymmetric systems.
In this case, $E_y$ is balanced by the bulk inertial term (the blue line)
at the X-line ($z = -0.2$).\citep{hesse14}
While traveling upward ($V_{ez}>0$),
the electrons are accelerated by $E_y$ in the $-y$ direction,
resulting in the bulk inertial effect.
The convection electric field (the red line) becomes zero
at the X-line and
at the flow stagnation point ($V_{ez}=0$) at $z = 0.4$. 
Above the X-line ($-0.2<z<0.2$),
the $\nabla\cdot\mathbb{P}_e$ term
mainly consists of the $\partial_z P_{eyz}$ term (the dotted line),
because the meandering electrons scatter their $y$-momentum in the ${\pm}z$ directions. 
Near the upper edge of the electron meandering region,
the $\partial_x P_{exy}$ term (the dashed line)
replaces the $\partial_z P_{eyz}$ term.
This is probably because
the rotation about $B_z$ carries away
the $y$-momentum in the ${\pm}x$ directions. 
Surprisingly,
the electron ideal condition is violated
below the X-line, $z \lesssim -0.3$. 
This was visible in previous studies,\citep{zeni11c,hesse14}
although they investigated other important issues. 
Around $-0.5 < z < -0.3$,
the negative $\nabla\cdot\mathbb{P}_e$ term cancels the bulk inertial term.
This signature was reported
on both upper and lower edges of the dissipation region
in symmetric reconnection,\citep{ishizawa05,klimas10}
but only the lower one is prominent in this asymmetric case. 
The $\partial_x P_{exy}$ term (the dashed line)
balances the nonideal electric field
below there, $z \lesssim -0.7$.
This signature has no counterpart in symmetric reconnection.

\subsection{Velocity distribution functions}
\label{sec:VDF}

\begin{figure*}[hthp]
\centering
\includegraphics[width={\textwidth},clip,bb=0 0 1135 543]{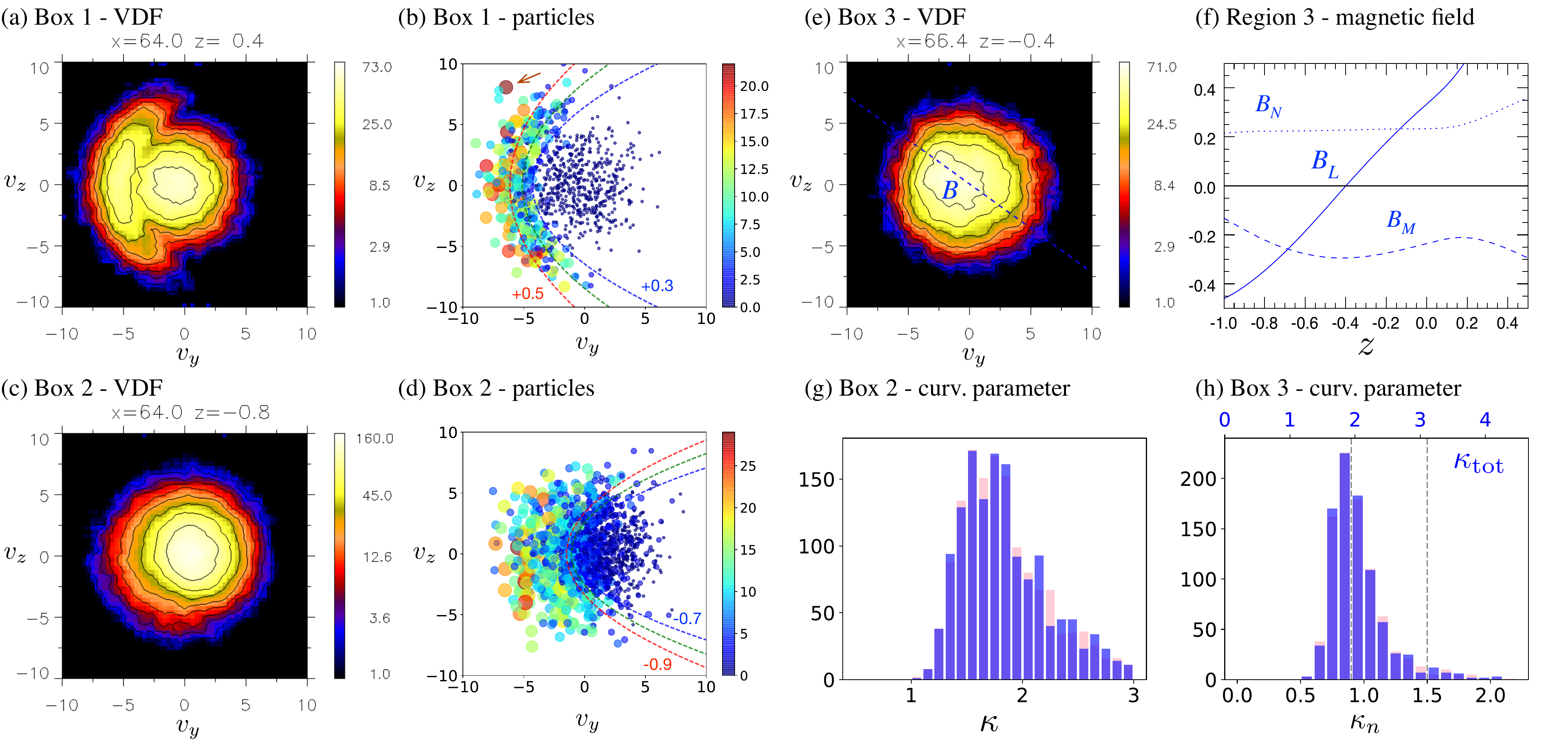}
\caption{(Color online)
\label{fig:VDF}
(a) The electron velocity distribution function
in $v_{y}$--$v_{z}$, integrated in Box 1.
(b) The velocity distribution of electrons in Box 1.
The color and size of the symbol stands for
the crossing numbers of the field reversal ($B_x=0$)
during $30<t<40$.
The dashed curves indicate conditions for the sheath-origin electrons
at $z=0.3$ (blue), $0.4$ (green), and $0.5$ (red).
See the text for more detail.
(c) The electron velocity distribution function
in $v_{y}$--$v_{z}$, integrated in Box 2.
(d) The velocity distribution of electrons in Box 2.
The dashed curves indicate conditions for the magnetospheric electrons
at $z=-0.7$ (blue), $-0.8$ (green), and $-0.9$ (red).
(e) The electron velocity distribution function
in $v_{y}$--$v_{z}$, integrated in Box 3.
(f) The magnetic fields in local $LMN$ coordinates
along the oblique line in Figure \ref{fig:snapshot}(c).
(g) The distribution of the electron curvature parameter $\kappa$ in Box 2
in an appropriately moving frame (blue) and in the simulation frame (red) at $t=35$.
(h) The distribution of the electron curvature parameters
$\kappa_{\rm tot}$ and $\kappa_n$ in Box 3.
The two dashed lines indicate parameters for Figure \ref{fig:poincare}.
}
\end{figure*}

Figures \ref{fig:VDF}(a), (c) and (e) display
the electron velocity distribution functions (VDFs),
accumulated in the three boxes in Figure \ref{fig:snapshot}.
The VDFs are shown in the $v_{y}$--$v_{z}$ plane,
integrated in the $v_{x}$ direction.

Figure \ref{fig:VDF}(a) shows the VDF in Box 1,
located in the magnetospheric side of the dissipation region.
As can be seen, it has
a crescent-shaped component in the left
and
a core component in the right.
The crescent stands for electron from the magnetosheath,
energized by the polarization electric field $E_z$ (Fig.~\ref{fig:snapshot}(b)).\citep{hesse14}
The core stands for magnetospheric electrons.
To confirm this scenario,
the electrons in the trajectory dataset are shown
in a scatter plot in Figure \ref{fig:VDF}(b).
We remind the readers that they are the three-percent subset of the total electrons.
Both the color and the size indicate
how many times the electrons cross the field reversal ($B_x=0$) during $30<t<40$.
As can be seen,
the electrons in the crescent part cross the field reversal many times,
while the electrons in the central core rarely cross the reversal.
The crossing numbers are higher in the outer part of the crescent,
because they gained their energy while traveling through the meandering orbit.

Considering the electron motion across the reconnection layer,
we find that the sheath-origin electrons should satisfy
the following inequality in the VDF, 
\begin{equation}
\label{eq:bessho}
v_{y} < 
-\Big(
\frac{1}{2e}mv_{z}^2
- \frac{1}{2m}eA_y^2(z)
- \phi(z)
\Big)
/
A_y(z)
\end{equation}
where $A_y(z) \equiv -\int_{z_0}^z B_x dz$ is the vector potential,
$\phi(z) \equiv -\int_{z_0}^z E_z dz$ is the scalar potential, and
the subscript $0$ denotes quantities at the X-line ($B_x=0$).
This inequality was originally proposed by \citet{bessho16}.
We will outline its derivation in Appendix \ref{sec:crescent}. 
The three dashed lines in Figure \ref{fig:VDF}(b) indicate
the conditions for the sheath-origin electrons (Eq.~\ref{eq:bessho}),
evaluated at the bottom ($z=0.3$; blue), middle ($z=0.4$; green),
and top ($z=0.5$; red) of Box 1.
The integrals $A_y(z)$ or $\phi(z)$
are calculated from the simulation data.
The blue curve excellently separates
the left population with a large number of crossings and
the central core population. 
By counting the crossing numbers during $30<t<35$,
we confirmed that
the electrons to the right of the blue curve originate from the magnetosphere
and have not crossed the reversal.
Most of the crescent electrons are sheath-origin (odd crossing numbers),
but some of them are sphere-origin (even crossing numbers).
They may be originally from the magnetosphere or
they may be temporally in the magnetosphere at $t=30$ during the meandering motion.
One can recognize in Figure \ref{fig:VDF}(b)
a few electrons to the right of the blue curve
do cross the field reversal.
They will enter the field-reversal regions sometime between $35<t<40$.

The influence of the normal electric field $E_z$ is found
in the last term in Eq.~\ref{eq:bessho}.
It is indicated by the green dashed curve in Figure \ref{fig:eohm}(a).
Around Box 1, the term is negative, and
it shifts the curves in Figure \ref{fig:VDF}(b) in the $-v_y$ direction.
The normal electric field $E_z$ energizes the meandering electrons,
shifting them in $-v_y$ in the velocity space around Region 1.
We also note that
the magnetospheric electrons are accelerated in $-y$
by the {\bf E}~$\times$~{\bf B} drift,
$V_{{\bf E}\times{\bf B},y} \approx E_z/B_x$
(the thin black curve in Figure \ref{fig:eohm}(a)).
However, 
the shift for the sheath-origin electrons is bigger than
the {\bf E}~$\times$~{\bf B} boost for the magnetospheric electrons,
$\phi(z)/A_y(z) = \int_{z_0}^z E_z dz ~\big/ \int_{z_0}^z B_x dz < E_z/B_x \approx V_{{\bf E}\times{\bf B},y}$,
because $E_z/B_x \approx V_{{\bf E}\times{\bf B},y}$ is
an increasing function from the negative infinity to zero.
The $\phi(z)/A_y(z) < V_{{\bf E}\times{\bf B},y}$ relation tells us that
the normal electric field $E_z$ acts to separate
the sheath-origin crescent electrons from the sphere-origin core electrons in the VDF.
The finite $E_z$ around the X-line makes the crescent pronounced
in the magnetospheric side. 

Figure \ref{fig:VDF}(c) shows the VDF in Box 2,
on the magnetosheath side of the X-line.
The particle view (Fig.~\ref{fig:VDF}(d)) reveals that
the VDF consists of a crescent-like component in the left
and a core component in the right.
Theoretically, Equation \ref{eq:bessho} is also applicable to the electrons in Box 2.
The three dashed lines indicate the inequality
at the top ($z=-0.7$; blue), middle ($z=-0.8$; green),
and bottom ($z=-0.9$; red) of the Box 2, respectively.
The crescent-like electrons cross the field reversal many times.
They correspond to the lower part of the meandering motion.
Electrons to the right of the blue curve
did not cross the reversal during $30<t<35$,
but they are going to cross the reversal after $t=35$.
Unlike the Box 1 case,
more electrons on the right side cross the reversal,
due to the {\bf E}~$\times$~{\bf B} upward motion. 
Even though they do not cross the reversal by $t=40$,
they will eventually reach the reversal sometime. 
Since the meandering electrons do not fully gyrate about the magnetic field line,
they often lead to violation of the electron ideal condition.
The meandering electrons hardly move in $z$ on average, and therefore
the electron bulk flow ${V}_{ez}$ becomes slower than the {\bf E}~$\times$~{\bf B} speed.
In such a case, we expect $E'_y \approx E_y + {V}_{ez} {B}_x = ({V}_{ez}-V_{{\bf E}\times{\bf B},z} ) {B}_x > 0$,
similar to the meandering region in the upper half of the dissipation region.
However, puzzlingly,
we observe the opposite sign $E'_y < 0$ around Region 2,
because the electron bulk flow outruns the {\bf E}~$\times$~{\bf B} flow,
${V}_{ez} > V_{{\bf E}\times{\bf B},z}$
(Figs.~\ref{fig:snapshot}(c) and \ref{fig:eohm}(b)).

\begin{figure}[htpb]
\centering
\includegraphics[width={\columnwidth},clip,bb=0 0 618 373]{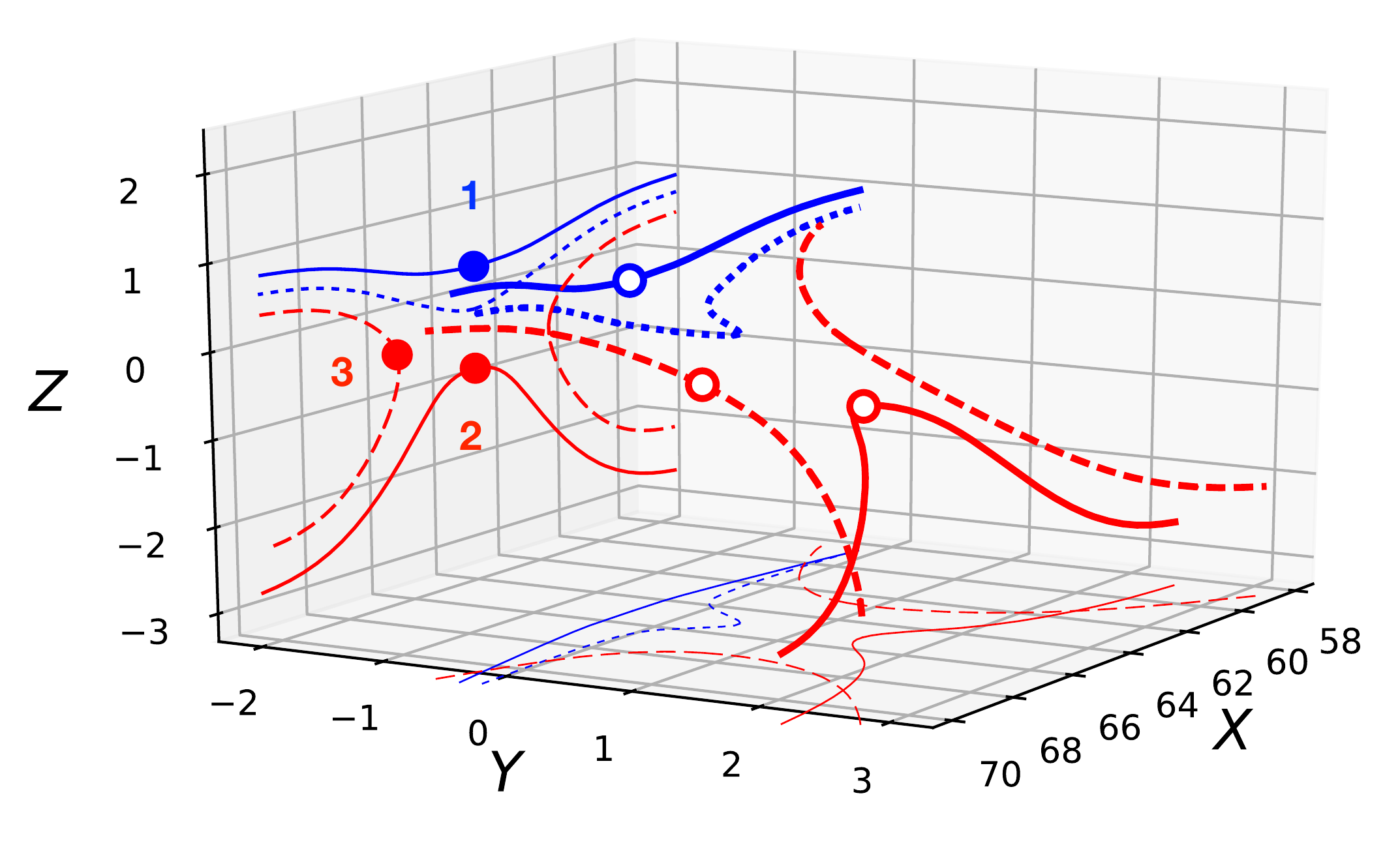}
\caption{(Color online)
\label{fig:3D}
The 3D structure and 2D projections of the magnetic field lines at $t=35$.
The circles correspond to the centers of Boxes 1, 2 and 3.
}
\end{figure}

Interestingly,
the ensemble curvature parameter is low ($\mathcal{K} \approx 2.2$) around Region 2
(Fig.~\ref{fig:snapshot}(f)).
This tells us that
the magnetic curvature radius is comparable with
the typical electron Larmor radius.
There are two reasons.
First, although the field lines look relatively straight in the 2D plane,
they are sharply bent in 3D around Region 2.
The 3D field lines and their 2D projections are presented in Figure \ref{fig:3D}.
The red solid line corresponds to
the field line through the center of Box 2.
It is stretched in the $y$ direction,
because the upward electron flow generates
the out-of-plane field $B_y$ (Fig.~\ref{fig:snapshot}(d)).
As a result, both $B_y$ and $B_z$ components reverse their signs
across the inflow line ($x=64.0$) around Region 2. 
The relevant magnetic curvature radius is $R_{\rm c} \approx 1.7$.
Second, the Larmor radius increases.
The magnetic field is $B_x=-0.31$ at $z=-0.8$ at the center of Box 2.
For an energetic electron of $|v_e|=5$, its Larmor radius is $\approx 0.65$,
which is comparable with the curvature radius $R_{\rm c}$.
We obtain $\kappa = 1.61$ in this case.
The electron Larmor radius further increases,
as it approaches the field reversal ($B_x=0$) at $z=-0.2$. 
In Figure \ref{fig:VDF}(g),
we present the distribution of curvature parameters
$\kappa$ for all the electrons in Box 2 at $t=35$. 
The curvature parameters in the blue histogram are calculated
in an appropriately moving frame
at the velocity of $(0.0, 0.33, 0.11) \approx \boldsymbol{V}_{{\bf E}\times{\bf B}}$.
We estimate this velocity such that
the electric energy density $\frac{1}{2}\varepsilon_0 |\boldsymbol{E}|^2$ in Box 2
becomes smallest.
Those in the red histogram are calculated
in the simulation frame.
The median values are $\kappa = 1.8$ in both cases.
One can see that most of electrons are
in the nonadiabatic regime of $\kappa \lesssim 2.5$.
We also note that
the ensemble curvature parameter $\mathcal{K} \approx 2.2$ is
a good indicator of the motion of these electrons.

In Figure \ref{fig:VDF}(e),
the dashed line indicates the direction of an average magnetic field in Box 3.
The VDF contains a field-aligned component,
streaming out from the magnetosheath side to the magnetospheric side.
We do not present a particle view,
because the field-reversal crossing is not meaningful,
as will be shown shortly. 
Figure \ref{fig:snapshot}(f) indicates that
the ensemble curvature parameter $\mathcal{K} \sim 1.6$ is the smallest
around the Box 3,
which is slightly above the $B_x$ reversal. 
This is due to the 3D field-line geometry. 
In addition to the in-plane field,
there is a guide field $B_y$ around Region 3. 
To better understand the topology,
we transform the coordinates into local $LMN$ coordinates.
We assume that the $N$ axis lies in the 2D simulation plane.
Then we find the $N$ axis by using the minimum variance method.\citep{sonnerup67}
It is indicated by the oblique line in Figure \ref{fig:snapshot}(c).
The $L$ direction is the maximum variance direction and
the $M$ direction completes the right-handed coordinate system.
The obtained $L$, $M$, and $N$ axes are similar to, but
slightly different from the $x$, $y$, and $z$ axes. 
Figure \ref{fig:VDF}(f) shows the magnetic field
along the $N$ axis, as a function of $z$.
Although the $B_x$ reversal is located at $z\approx -0.6$,
the $B_L$ reversal is located at $z = -0.4$. 
This corresponds to the local minimum of $\mathcal{K}$.
Using the shear-field model in Section \ref{sec:kappa},
we evaluate the curvature parameters
$\kappa_n$ and $\kappa_{\rm tot}$ (Eq.~\ref{eq:BZ91})
for all the electrons in Box 3 at $t=35$.
Based on the $LMN$ analysis,
the shear field is set to $B_s/B_n = -1.25$.
The histograms in Figure \ref{fig:VDF}(h) present
the distribution of the electron curvature parameters.
The frame-transform velocity $(0.24,0.08,-0.04)$ for the blue histogram
is virtually negligible.
The parameters are
in the range of $0.5 < \kappa_n < 1.5$ and $1.0 < \kappa_{\rm tot} < 3.0$.
The electrons are basically in the nonadiabatic regime. 
Their median value $\kappa_{\rm tot} = 1.8$ is well represented by
the ensemble curvature parameter $\mathcal{K}\sim 1.6$.
Note that $\mathcal{K}\sim 1.6$ is the local minimum along the oblique line.
It has some variations even in Box 3. 
We have also confirmed that
the electrons are in the nonadiabatic regime
by examining Poincar\'{e} sections.\citep{BZ91}
The analysis is summarized in Appendix \ref{sec:poincare}.

\subsection{Particle orbits}
\label{sec:orbit}

\begin{figure*}[htpb]
\centering
\includegraphics[width={\textwidth},clip,bb=0 0 807 1003]{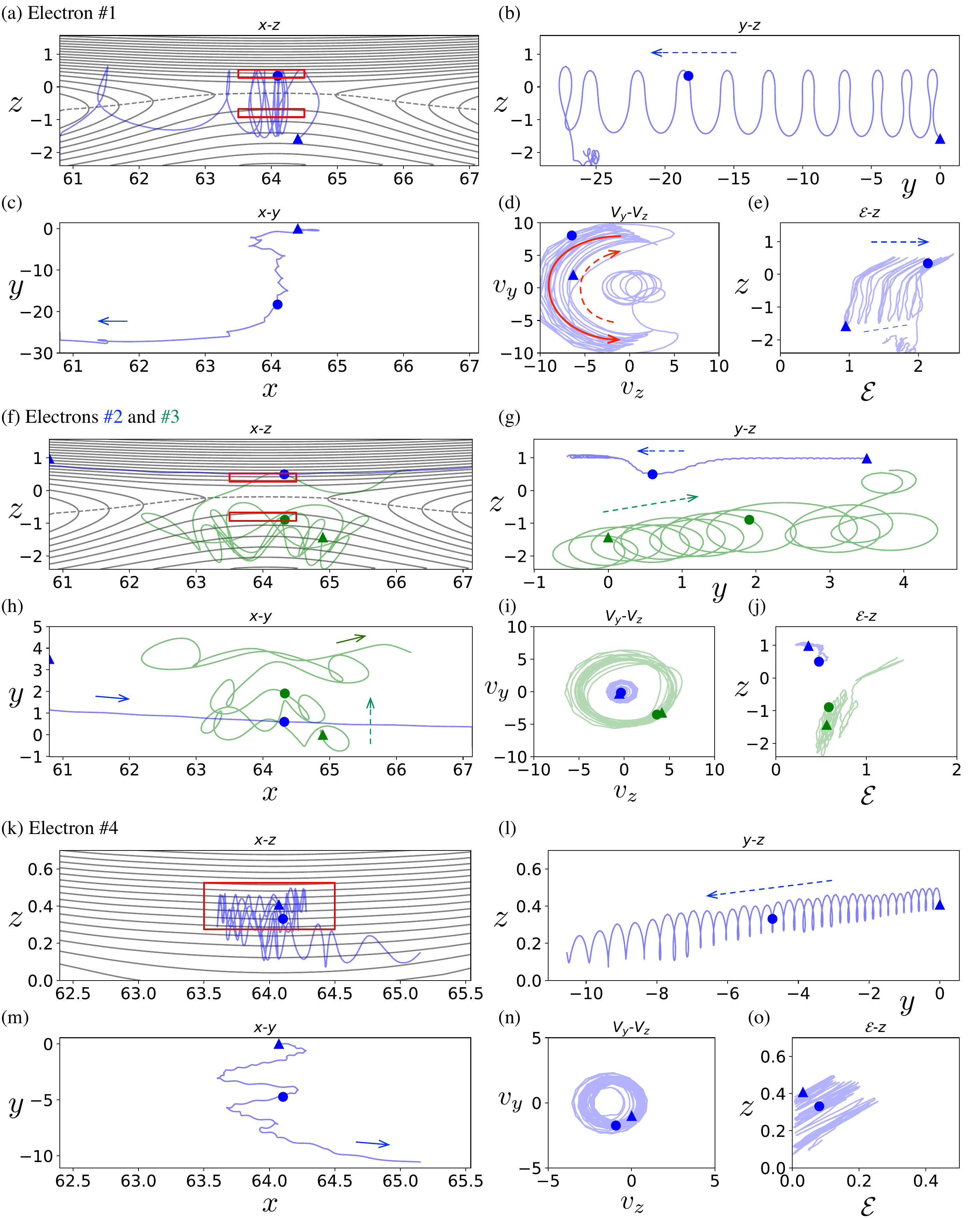}
\caption{(Color online)
\label{fig:orbit}
The electron orbits during $30<t<40$
in the $x$--$z$ (a,f,k), $y$--$z$ (b,g,l) and $x$--$y$ (c,h,m) planes,
in the velocity ($v_y$--$v_z$) space (d,i,n), and
in the energy ($\mathcal{E}$--$z$) space (e,j,o).
The triangles and circles indicate
the positions at $t=30$ and at $t=35$, respectively. 
The red boxes stand for Boxes 1 and 2.
}
\end{figure*}

In the trajectory dataset,
1845 electrons are located in Boxes 1--3 at t=35.
We visually inspected 600 of them and then
we show representative ones in Figure \ref{fig:orbit}. 
The first three panels show the orbits
in the $x$--$z$, $y$--$z$, and $x$--$y$ planes.
The next two panels present
the orbits in the velocity space ($v_y$--$v_z$) and
in the energy space ($\mathcal{E}$--$z$).
The triangles and the circles indicate
the positions at $t=30$ and $t=35$, respectively.

The first case represents
a meandering orbit (Figs.~\ref{fig:orbit}(a)--(c)).
Starting from the magnetosheath,
the electron continues to bounce around the center.
It travels a long distance in $-y$,
crossing the field reversal 22 times.
In the velocity space (Fig.~\ref{fig:orbit}(d)),
it moves along the left arcs. 
Obviously the electron belongs to
the crescent-shaped component in the VDF (Fig.~\ref{fig:VDF}(a)).
The electron at $t=35$ is indicated by the small arrow in Figure \ref{fig:VDF}(b).
When the electron is above the field reversal ($z>-0.2$),
it goes around the outer side of the crescent,
as indicated by the red arrow in Figure~\ref{fig:orbit}(d). 
Below the reversal, it slowly moves
in the inner side of the crescent (the dashed arrow). 
This is because
the electron gains substantial energy from negative $E_z$ (Fig.~\ref{fig:snapshot}(b)).
This is evident in the energy diagram in Figure~\ref{fig:orbit}(e).
Despite the lower energy,
it travels much deeper in $z$ in the magnetosheath,
because $|B|$ is weaker there.
During the meandering motion,
the electron continuously gains the energy from the reconnection electric field. 
Figure~\ref{fig:orbit}(e) indicates that
the electron doubled its energy from $t=30$ to $t=40$. 
It shifts to $-v_y$ in the velocity space (Fig.~\ref{fig:orbit}(d)).
The meandering length in $y$ gradually increases accordingly.
Meanwhile,
the meandering width in $z$ slowly decreases from $-1.55<z<0.535$ to $-1.4<z<0.5$,
when the electron stays around the center
($x>63.5, y>-24.3$; Figs.~\ref{fig:orbit}(a) and (b)).
One can see this $z$-confinement
in the lower envelope of the orbit
in Figure \ref{fig:orbit}(e).
The $z$-confinement qualitatively agrees with
a damped oscillation during the Speiser orbit.\citep{speiser65}
It is more evident in lower-energy meandering electrons,
because they are more sensitive to
the incoming {\bf E}~$\times$~{\bf B} flows than high-energy electrons.
Concerning the particle energization,
since the normal electric field ($|E_z|$) is
an order-of-magnitude stronger than the reconnection electric field $|E_y|$,
the normal field is a particle accelerator on a short time-scale.
On the other hand,
the $y$-acceleration continues as long as the electron runs in $-y$.
Figure~\ref{fig:orbit}(e) shows that
the electron gained more energy from $E_y$ than from $E_z$. 
Therefore, the reconnection electric field $E_y$ is a major accelerator
on a long time-scale for the highest-energy electrons. 

In the middle panels,
the electron \#2 in blue corresponds to
an electron in the core component in the VDF in Box 1
(Fig.~\ref{fig:VDF}(a)).
It has a high $x$-velocity, $v_{x} \approx 4.9$ at $t=35$.
The electron starts from $x=42$, far outside the presented domain, at $t=30$.
Then, it passes Box 1 along the field lines. 
The electron drifts in the $-y$ direction (Fig.~\ref{fig:orbit}(g)),
due to $E_z$ around the separatrix boundary. 

The electron \#3 in green stays
in the magnetosheath for a long time.
While bouncing in ${\pm}x$,
it slowly moves in the $(+y, +z)$ direction,
as indicated by the green dashed lines (Figs.~\ref{fig:orbit}(g,h)).
Judging from the displacement of its guiding center (Fig.~\ref{fig:orbit}(g)),
the average drift speed is $\sim (0.0, 0.4, 0.16)$.
The {\bf E}~$\times$~{\bf B} velocity around Box 2 is
$\sim (0.0, 0.3, 0.1)$ (Fig.~\ref{fig:eohm}(a)). 
On average, the gradient-{\bf B} drift probably explains
the additional drift in $+y$.
Later, it crosses the field reversal and then
escapes to the magnetospheric side in the $+x$ direction. 
When it jumps to the magnetospheric side,
the electron \#3 gains substantial energy by $E_z$,
as evident in Figure \ref{fig:orbit}(j).
It is located to the right of the blue curve
in Figure \ref{fig:VDF}(d), and
it did not cross the reversal at $t=35$.
After some time, it eventually escapes to the magnetospheric side.

In the bottom panels,
the electron \#4 stays around Region 1.
It drifts in the ($-y$, $-z$) direction
due to the normal electric field $E_z$ and
the reconnection electric field $E_y$.
The electron bounces in $x$,
because it is trapped by $E_{\parallel}$ (Fig.~\ref{fig:snapshot}(d))
as discussed by \citet{egedal11}. 
This and the electron \#2 belong to the core component
in the $v_{y}$--$v_{z}$ space.
However, unlike the electron \#2,
the electron \#4 has a lower $x$-velocity ($|v_{x}| \le 1$) and
therefore it is trapped there. 
Later, the electron \#4 escapes in the $+x$ direction,
because $E_{\parallel}<0$ at $x\approx 65$
(the blue region Fig.~\ref{fig:snapshot}(d)). 
Other electrons may eventually reach the field reversal and then
start to undergo the meandering motion, however,
it will take a longer time to reach the field reversal at $z_0=-0.2$.
Although it is almost impossible to see in Figure \ref{fig:eohm}(b),
the {\bf E}~$\times$~{\bf B} speed is
$V_{{\bf E}\times{\bf B},z} \sim -0.025$ at $z>0.2$. 
The guiding center moves only $\Delta z \sim 0.125$ during $35<t<40$,
as seen in Figure \ref{fig:orbit}(l).

\subsection{Electron mixing}
\label{sec:mix}

We propose a simple parameter to evaluate
the electron mixing during magnetic reconnection. 
ince magnetic reconnection mixes plasmas from two inflow regions,
we classify the electrons,
based on the polarity of $B_x$ at the particle position at $t = t_0-\Delta t$.
Here, we employ the magnetic field at the particle position rather than
at the guiding center,
because the guiding center may not be meaningful for unmagnetized particles.
Then, to quantify the mixing of two populations,
we define a forward-time {\it finite-time mixing fraction} (FTMF),
\begin{align}
\label{eq:M_in}
\mathcal{M}_{\rm f}(t_0,\Delta t)
\equiv {\rm mix}
\big(
& N \big[ B_x(\boldsymbol{r}(t_0-\Delta t), t_0-\Delta t)< 0 \big], \nonumber\\
& N \big[ B_x(\boldsymbol{r}(t_0-\Delta t), t_0-\Delta t)> 0 \big]
\big)
,
\end{align}
where ${\rm mix}( N_1,N_2 ) \equiv {2\min ( N_1,N_2 )}/{(N_1+N_2)}$ is a mixing function,
$N[~]$ is the number of electrons that satisfy the condition in the square brackets,
$B_x(\boldsymbol{r},t)$ is the $x$ component of magnetic field
at the position $\boldsymbol{r}$ and time $t$,
and $\boldsymbol{r}(t)$ is the position vector of individual electrons at $t$.
The mixing function returns a value
between 0 ($N_1=0$ or $N_2=0$, no mixing) and 1 ($N_1=N_2$, well mixed).

Furthermore, magnetic reconnection is something more than a simple mixing process.
It ejects plasmas in the outflow directions.
Recognizing that
this is a mixing process from the outflow regions in the {\it reverse-time} direction,
we define a backward-time FTMF
by using the polarity of $B_z$ at the future position at $t=t_0+\Delta t$,
\begin{align}
\label{eq:M_out}
\mathcal{M}_{\rm b}(t_0,\Delta t)
\equiv {\rm mix}
\big(
& N \big[ B_z(\boldsymbol{r}(t_0+\Delta t), t_0+\Delta t)< 0 \big], \nonumber\\
& N \big[ B_z(\boldsymbol{r}(t_0+\Delta t), t_0+\Delta t)> 0 \big]
\big)
.
\end{align}
Combining the forward-time and backward-time FTMFs,
we introduce a mixing measure for reconnection,
\begin{equation}
\label{eq:M}
\mathcal{M}_R(t_0,\Delta t)
\equiv
\mathcal{M}_{\rm f}(t_0,\Delta t)
\times
\mathcal{M}_{\rm b}(t_0,\Delta t)
.
\end{equation}
We call it the FTMF or the mixing parameter. 
Since both $\mathcal{M}_{\rm f}$ and $\mathcal{M}_{\rm b}$ range from 0 to 1,
$\mathcal{M}_R \approx 1$ characterizes
a key region surrounding the X-line,
where the two inflow populations mix with each other and then
start to escape in the two outflow directions. 
The idea to use forward- and backward-time mixing parameters comes from
a new diagnosis in geophysical fluid dynamics (e.g. \citet{haller15}),
in which forward- and backward-time finite-time Lyapunov exponents (FTLEs)
visualize flow boundaries.

To evaluate Equation \ref{eq:M},
we need three footprint data at $t=t_0$ and $t=t_0 \pm \Delta t$.
One can obtain a similar result
by evaluating
$\mathcal{M}_R(t_0,\Delta t)
\approx
\mathcal{M}_{\rm f}(t_0+\frac{1}{2}\Delta t,\Delta t)
\times
\mathcal{M}_{\rm b}(t_0-\frac{1}{2}\Delta t,\Delta t)$
from two footprints.
We employ the former approach in this study,
because it preserves an original meaning that
particles mix and then they depart.
Using the trajectory dataset,
we evaluate $\mathcal{M}_R(35,\Delta t)$
for the range of $\Delta t = 0 \rightarrow 5$.
The electron circulation effects across the $x$-boundaries are ruled out.
During $\Delta t=5$, the highest-energy electron
($|v| \approx 10$; see Fig.~\ref{fig:orbit}d for example)
may travel $\lesssim 50 ~d_i$,
but the $x$-boundaries are $57 ~d_i$ away from
the domain in Figure \ref{fig:snapshot}(h). 
Then, by inspecting a series of the results,
we choose $\Delta t = 1.0~\Omega_{\rm ci}^{-1} = 25.0~\Omega_{\rm ce}^{-1}$.
The resulting FTMFs
$\mathcal{M}_R(35,1)$, $\mathcal{M}_f(35,1)$, and $\mathcal{M}_b(35,1)$
are presented in Figures \ref{fig:snapshot}(g) and \ref{fig:snapshot}(h).
We use coarse cells,
because only three percent of full particle data are available from the trajectory dataset.
Each cell contains $50$--$200$ electrons. 

In Figure \ref{fig:snapshot}(h),
$\mathcal{M}_f(35,1)$ in red tells us that
the upper-origin and lower-origin electrons are well mixed
inside the entire current layer.
In addition to the mixing in the reconnection dissipation region,
the reconnected magnetic field allows electron mixing inside the exhaust region.
On the other hand,
$\mathcal{M}_b(35,1)$ in blue
marks the inflow regions and the X-line vicinity.
This captures the bounce motion in the inflow regions.\citep{egedal11}

In Figure \ref{fig:snapshot}(g),
the FTMF $\mathcal{M}_R$ marks several places.
In particular, it emphasizes the magnetospheric side of the X-line.
As guided by the red dashed box in Figure \ref{fig:snapshot},
the length in $x$ and the width in $z$ are comparable with
the energy-dissipation region identified by $\mathcal{D}_e$.
There, the sheath-origin crescent electrons and the sphere-origin core electrons
coexist in the VDF (Figs.~\ref{fig:VDF}(a,b)). 
Then two populations are both ejected to either of the two outflow regions.
These behaviors are captured by $\mathcal{M}_f$ and $\mathcal{M}_b$
(Eqs.~\ref{eq:M_in} and \ref{eq:M_out}). 
A careful inspection tells us that
$\mathcal{M}_R$ marks T-shaped region in the red box.
This is probably because
the electrons near the bottom-left and bottom-right corners 
are threaded by the outermost magnetic field lines in ${\pm}x$. 
The region between Regions 1 and 2 is weakly marked by $\mathcal{M}_R$,
because the electrons are meandering in ${\pm}z$.
However, $\mathcal{M}_R$ is remarkably smaller in Region 2 than in Region 1.
First, because of the density asymmetry,
the VDF in Box 2 (Figs.~\ref{fig:VDF}(c,d)) contains
much more electrons in the core component than in Box 1.
As shown in Section \ref{sec:orbit} (see the orbit \#3 in Fig.~\ref{fig:orbit}),
they spent a relatively long time in the magnetosheath
and therefore
they do not cross the field reversal on a short timescale of $\Delta t = 1.0$.
Therefore, the number of meandering electrons is relatively small in Region 2.
Second, as already discussed,
the meandering electrons have lower energies in the magnetosheath side
than in the magnetospheric side (Fig.~\ref{fig:orbit}(e)).
Since they move slower in the magnetosheath, the electron mixing is less efficient.
For these reasons, $\mathcal{M}_R$ is small around Region 2.

One can also see the four arms along the separatrices,
because the electrons quickly move
to and from the central mixing site along the field lines. 
The FTMF region stretches to the outward directions, and
it will eventually spread over the entire current layer
in the $\Delta t \rightarrow \infty$ limit.
Therefore, it is important to choose an appropriate $\Delta t$
that represents the electron kinetic physics around the X-line. 
There are weak mixing sites in the exhaust regions at $x<60$ and $x>68$.
They correspond to the magnetic O-type regions. 
Some electrons travel along the field lines inside the magnetic islands,
and then they repeatedly go through
the $B_x<0$ and $B_x>0$ regions and the $B_z<0$ and $B_z>0$ regions. 
Therefore the FTMF occasionally detects such O-type magnetic islands.

\section{Discussion}
\label{sec:discussion}

We have examined several aspects of electron kinetic physics
during asymmetric magnetic reconnection in an anti-parallel magnetic field. 
Due to the asymmetric plasma conditions,
the electron physics appears in very different ways
in the two nonideal regions across the reconnection layer (Regions 1 and 2).

On the magnetospheric side of the X-line (Region 1),
the electron momentum transport in
the pressure tensor term, $\nabla\cdot\mathbb{P}_e$,
balances the reconnection electric field
at the electron stagnation point.\citep{hesse14}
By tracking self-consistent electron trajectories,
we have verified recent results that
the VDF consists of
the crescent-shaped component by the sheath-origin meandering electrons
and
the core component by magnetospheric electrons.\citep{hesse14,bessho16,chen16,shay16,egedal16}
Examining the one-dimensional theory,
we have further found that the normal electric field around the X-line
makes the crescent pronounced. 
Although Region 1 is not marked by the curvature parameters,
the crescent part of the VDF is obviously nonadiabatic.

Since the electron orbits are fundamental elements
for the kinetic physics of magnetic reconnection,
it is interesting to see the relevance to the orbits in symmetric cases.
In a symmetric system,
\citet{zeni16} recently found many ``noncrossing electrons''
that do not cross the field reversal, due to the Hall electric field $E_z$. 
In this study, in Figure~\ref{fig:VDF}(b),
most of the core electrons in Box 1 do not cross the field reversal,
because they are kept away from the reversal by $E_z$ along the sphere-side separatrix. 
They correspond to the noncrossing electrons in the symmetric case. 
\citet{zeni16} further found
Speiser-type noncrossing orbits in the symmetric system.
In this asymmetric case,
we fail to distinguish the noncrossing Speiser-like orbits,
because the magnetic topology is very flat in the magnetospheric side.
In the magnetosheath,
we do not expect noncrossing electrons
because there is no electric field $E_z$ pointing toward the field reversal.
Some electrons do not cross the field reversal
in the VDF in Box 2 (Fig.~\ref{fig:VDF}(d)). 
However, as discussed,
we do not consider them as noncrossing electrons,
because they will eventually cross the field reversal after $t>40$.

We have examined electron-physics signatures in Region 2 in the magnetosheath side.
The curvature parameters suggest that
the electron are in the nonadiabatic regime.
The electron motion becomes chaotic, because
the electron Larmor radius is comparable with the magnetic curvature radius.
The curvature is actually steep in 3D, and
the Larmor radius is large in a weak magnetic field. 
A typical nonadiabatic motion is shown
in the orbit \#3 in Figure \ref{fig:orbit}(f).
Since the nonadiabatic electrons no longer follow
the magnetized drift motion,
the ideal condition need not be preserved. 
Strictly speaking, there is no guarantee that
the nonadiabatic electrons always violate the ideal condition.
However, the system requires the electric current to maintain the magnetic curvature.
Only electrons can carry the currents for small-scale magnetic curves,
while the ions are insensitive to the electron-scale structure.
In reality, nonadiabatic electrons do carry a huge amount of electric current.
Assuming $-en\boldsymbol{V}'_e \approx \boldsymbol{J}$
in the {\bf E}~$\times$~{\bf B} frame,
we expect $\boldsymbol{E}+\boldsymbol{V}_e \times \boldsymbol{B} \approx -\frac{1}{en}\boldsymbol{J}\times\boldsymbol{B}$.
The ideal condition is violated in the $-\boldsymbol{J}\times\boldsymbol{B}$ direction.
Around Region 2, the $y$ component along the inflow line yields
$E'_y = [ \boldsymbol{E}+\boldsymbol{V}_e \times \boldsymbol{B} ]_y \approx -\frac{1}{en\mu_0} B_x \partial_x B_y < 0$, 
in agreement with Figure \ref{fig:eohm}(d).
The electron Ohm's law is supported by the ${\nabla \cdot} \mathbb{P}_e$ term at $z=-0.8$,
because the inertial term is still small.
We argue that Region 2 is a vertical variant of
a nongyrotropic current layer in symmetric reconnection.\citep{hesse08,zeni16}
If we consider a one-dimensional current layer,
$\boldsymbol{J}=\boldsymbol{J}(x)$ and $\partial_z\approx 0$,
it is reasonable to see that
the pressure tensor term is supported by the variation in $x$,
$({\nabla \cdot} \mathbb{P}_e)_y \approx \partial_x P_{exy}$ (Fig.~\ref{fig:eohm}(b))
at $z \lesssim -0.8$.
In Section \ref{sec:results},
we have reported the puzzling signatures around Region 2,
such as $\boldsymbol{V}_{e} \ne \boldsymbol{V}_{{\bf E}\times{\bf B}}$ and ${E}'_y \ne 0$. 
They are attributed to the nonadiabatic behavior of electrons. 
The electron nonidealness
in the magnetosheath side of the X-line
is consistent with previous PIC simulations
(see Fig.~2b of \citet{prit09a}, Fig.~11b of \citet{mozer11},
Fig.~3 of \citet{hesse14}, and Fig.~2g of \citet{lea17}). 

Around Region 3,
we have found an electron nonideal layer.
As presented in Figures \ref{fig:snapshot}(f) and \ref{fig:VDF}(f),
the nonadiabatic (nonideal) layer corresponds to the field reversal $B_L=0$
in appropriately rotated coordinates.
We argue that the nonidealness stems from
the nonadiabatic motion of electrons.
The layer resembles a fast electron-jet inside the exhaust
in symmetric reconnection.\citep{kari07,shay07,hesse08}
The symmetric electron-jet is populated by Speiser-orbit electrons,\citep{zeni16}
as characterized by $\mathcal{K} < 1$.\citep{lea13}
In this asymmetric case,
electrons are not in the Speiser regime, but
in the nonadiabatic regime of $1 < \mathcal{K} \lesssim 2.5$.
In both symmetric and asymmetric cases,
the electron nonideal layer is virtually non-dissipative,
as seen in Figure~\ref{fig:snapshot}(e).
The transition from the symmetric $\mathcal{K} < 1$ layer to
the asymmetric $1 < \mathcal{K} \lesssim 2.5$ layer
would be sensitive to the mass ratio, the guide field, and
the asymmetry in inflow parameters,
in analogy with symmetric systems.\citep{lea13}

Let us estimate whether we will see the nonadiabatic signatures in the actual world. 
We assume that
the magnetic field topology is determined by
a hybrid scale of the electron and ion physics. 
Then we expect
the minimum magnetic curvature radius to be
a geometric mean of the local inertial lengths of ions and electrons,
$R_{\rm c,min} \sim (d_i d_e)^{1/2}$. 
The ensemble curvature parameter (Eq.~\ref{eq:K}) yields
\begin{equation}
\label{eq:estimate}
\mathcal{K} =
\Big( \frac{R_{\rm c,min}}{\rho_{\rm eff}} \Big)^{1/2}
\sim
\Big( \frac{d_{i}d_{e}}{\rho^2_{\rm eff}} \Big)^{1/4}
=
\Big(\frac{2}{\beta_e}\Big)^{1/4}
\Big(\frac{m_i}{m_e}\Big)^{1/8}
,
\end{equation}
where $\beta_e$ is the electron plasma $\beta$. 
Our initial conditions $m_i/m_e=25$ and $\beta_e=2$ in the sheath side
give $\mathcal{K} \sim 1.5$,
in agreement with $\mathcal{K} \lesssim 2.2$ above Region 2 (Fig.~\ref{fig:snapshot}(f)).
If we assume $\mathcal{K} < 2.5$ for nonadiabatic signatures,
the magnetosheath plasma $\beta$ needs to be
$\beta > 13.1$ for $T_i = 5T_e$ and $\beta > 4.4$ for $T_i = T_e$
at the real mass ratio. 
This is possible in the magnetosheath.
The plasma $\beta$ becomes even an order-of-magnitude higher
inside the reconnection outflow exhaust.
Therefore, we expect the nonadiabatic signatures,
at the sheath-side vicinity of the X-line and
inside the outflow region near the X-line at the dayside magnetopause.
Along with PIC simulations at higher mass ratios,
the nonadiabatic signatures
could be observed near the X-line with MMS. 

Recently, \citet{hwang17} observed
the violation of the electron ideal condition
in the exhaust region in magnetopause reconnection with MMS.
Their result corresponds to $E'_y<0$ in our coordinates.
During the event,
the out-of-plane magnetic field was finite
around the field reversal (Fig.~2(a) of \citet{hwang17}).
The relevant electron VDF contained a parallel component,
streaming away from the magnetosheath (Figs.~3(e) and 2(n) of \citet{hwang17}).
All these results are consistent with our prediction for Region 3. 
Unfortunately, the curvature parameter was not clear in this event. 
Using the minimum magnetic field 10 nT and the typical electron energy 100 eV,
we estimate the Larmor radius to $\rho_{\rm eff} = 3$--$4$ km. 
In the nonadiabatic case of $\mathcal{K} < 2.5$,
the magnetic curvature radius was supposed to be
$R_c = \mathcal{K}^2 \rho_{\rm eff} < 18$--$25$ km.
This is smaller than
the average spacecraft separation 64 km at that time. 
Fortunately, the separation often remained about $10$ km or below
during the first phase of the mission.
Such separations should be sufficient to confirm the curvature in similar events.

We have also proposed the FTMF to evaluate the electron mixing.
Previous studies on the plasma mixing
during Kelvin-Helmholtz instability\citep{yosuke06,tkm11,delamere11} and
magnetic reconnection\citep{dau14,zeni16}
employed the following or similar parameters,
\begin{eqnarray}
\label{eq:M_in2}
\mathcal{R}_{\rm mix}(t_s,t_0)
&\equiv &
{\rm mix}
\big(
N \big[ \boldsymbol{r}(t_s)_z > 0 \big],
N \big[ \boldsymbol{r}(t_s)_z < 0 \big]
\big)
,
\\
\label{eq:F}
\mathcal{F}(t_s,t_0)
&\equiv &
\frac{N \big[ \boldsymbol{r}(t_s)_z > 0 \big]-N \big[ \boldsymbol{r}(t_s)_z < 0 \big]}{N \big[ \boldsymbol{r}(t_s)_z > 0 \big]+N \big[ \boldsymbol{r}(t_s)_z < 0 \big]}
= \pm \big[ 1 - \mathcal{R}_{\rm mix}(t_s,t_0) \big]
,
\end{eqnarray}
where $t=t_s$ is the start time.
We have extended these diagnoses
by combining the two mixing fractions and
by using a finite mixing time $\Delta t$ in the start time, $t_s=t \pm \Delta t$. 
The best value for $\Delta t$ is under investigation.
For electrons, the mixing time should be determined by
a typical timescale of the electron physics. 
Meanwhile, to distinguish reconnection-physics phenomena from a gyration,
it should be longer than the electron gyroperiod
$\Delta t > 2\pi\Omega_{\rm ce}^{-1}$.
Our current choice of $\Delta t = 25.0~\Omega_{\rm ce}^{-1}$ satisfies
these two conditions.
In order to classify
the two inflow populations for $\mathcal{M}_{\rm f}$ and
the outflow populations for $\mathcal{M}_{\rm b}$,
we employed $B_x$ and $B_z$,
because we knew that the initial magnetic field is in ${\pm}x$
and that the reconnection occurs in the $x$--$z$ plane.
However, it is not certain how to extend these methods
to generic cases in 2D and 3D.
Numerical tests at various mass ratios in various configurations are necessary,
to find the best mixing time $\Delta t$ and
the best conditions for $\mathcal{M}_{\rm f}$ and $\mathcal{M}_{\rm b}$.

Physically, it is very interesting that
the mixing site is similar to the energy dissipation site, 
identified by $\mathcal{D}_e$ (Fig.~\ref{fig:snapshot}(e)).
As shown in Figures~\ref{fig:VDF}(a,b),
two populations of different origins coexist in the VDF. 
Then they will mix with each other in the phase space,
as they escape in the outflow directions.
We expect that
such a phase-space mixing involves
the local electron heating and the entropy increase. 
This should involve the plasma heating in the local MHD frame
or the nonideal energy transfer.\citep{zeni11c}
The relationship among
the electron mixing, the plasma heating, and the entropy evolution
deserves further investigation. 
This is the first step to quantitatively evaluate
the electron mixing during magnetic reconnection.

\section{Conclusions}
\label{sec:conclusion}

We have studied several properties of electron kinetic physics
during asymmetric magnetic reconnection in an anti-parallel configuration,
by using the 2D PIC simulation.
We have focused on three characteristic regions near the X-line,
where the electron ideal condition is violated.
On the magnetospheric side of the X-line,
the normal electric field enhances
the electron meandering motion from the magnetosheath.
The motion leads to a crescent-shaped component in the VDF,
in agreement with previous studies. 
On the magnetosheath side,
since the magnetic field line is stretched in the third dimension and
since the magnetic field is weak,
the magnetic curvature radius is comparable with the electron Larmor radius. 
The electron motion becomes highly nonadiabatic, and therefore
the electron idealness is no longer expected to hold. 
Around the middle of the outflow regions,
the electron nonidealness is coincident with
the region of the nonadiabatic motion.
These nonadiabatic signatures would be observable
at the actual magnetopause.
Utilizing the PIC data,
we have introduced the FTMF to diagnose the electron mixing.
We have found that
the electron mixing is enhanced on the magnetospheric side of the X-line,
where the nonideal energy dissipation occurs.
This suggests that
the electron mixing produces the plasma heating and
the nonideal energy dissipation
during magnetic reconnection.

\begin{acknowledgments}
The authors acknowledge T. K. M. Nakamura for helpful comments.
This work was supported by the facilities at Center for Computational Astrophysics,
National Astronomical Observatory of Japan and
the Information Technology Center, Nagoya University.
Because of the large storage requirements,
the simulation data are not publicly available.
Interested researchers are welcome to contact the first author.
This work was supported by
Grant-in-Aid for Scientific Research (C) 17K05673
from the Japan Society for the Promotion of Science (JSPS).
\end{acknowledgments}

\appendix

\section{1D crescent model}
\label{sec:crescent}

Here we outline \citet{bessho16}'s inequality for the VDF,
taking recent advances into account.\citep{shay16,egedal16,lapenta17}
We consider a quasi-static 1D reconnection layer along the inflow line.
The magnetic and electric fields are approximated by
$\boldsymbol{B}(z) = ( B_x(z), 0, 0 )$ and $\boldsymbol{E}(z) = ( 0, 0, E_z(z) )$.
Here we neglect $E_y$, because $E_z$ is the strongest component.
The vector and electrostatic potentials satisfy
$B_x = -\partial_z A_y$ and $E_z = -\partial_z\phi$.
The conservation of the canonical momentum and the energy yields
\begin{eqnarray}
v_{y0}
&=&
v_{y}
- \frac{e}{m} A_y(z)
\label{eq:py}
\\
\frac{1}{2}(v_{y0}^2+v_{z0}^2)
&=&
\frac{1}{2}(v_{y}^2+v_{z}^2)
- \frac{e}{m} \phi(z)
\label{eq:ene}
\end{eqnarray}
where
the subscript $0$ denotes quantities at the X-line ($B_x=0$),
$A_y(z) \equiv -\int_{z_0}^z B_x dz$ is the vector potential, and
$\phi(z) \equiv -\int_{z_0}^z E_z dz$ is the scalar potential. 
The reference point of the potentials is set to the X-line. 
Substituting Eq.~\eqref{eq:py} into Eq.~\eqref{eq:ene}, we obtain
\begin{equation}
\frac{1}{2}v_{z0}^2 =
\frac{1}{2}v_{z}^2
+ v_y\frac{e}{m} A_y(z)
- \frac{1}{2}\Big[ \frac{e}{m} A_y(z) \Big]^2
- \frac{e}{m}\phi(z)
.
\end{equation}
If the electron reaches $z=z_0$, $v_{z0}^2>0$ needs to be satisfied.
This leads to the inequality (Eq.~\ref{eq:bessho} in Sec.~\ref{sec:results}),
\begin{equation}
v_{y} < 
-\Big(
\frac{1}{2e}mv_{z}^2
- \frac{1}{2m}eA_y^2(z)
- \phi(z)
\Big)
/
A_y(z)
.
\end{equation}
This form allows arbitrary profiles of $B_x(z)$ and $E_z(z)$.

\section{Poincar\'{e} map}
\label{sec:poincare}

\begin{figure}[htpb]
\centering
\includegraphics[width={\columnwidth},clip,bb=0 0 558 283]{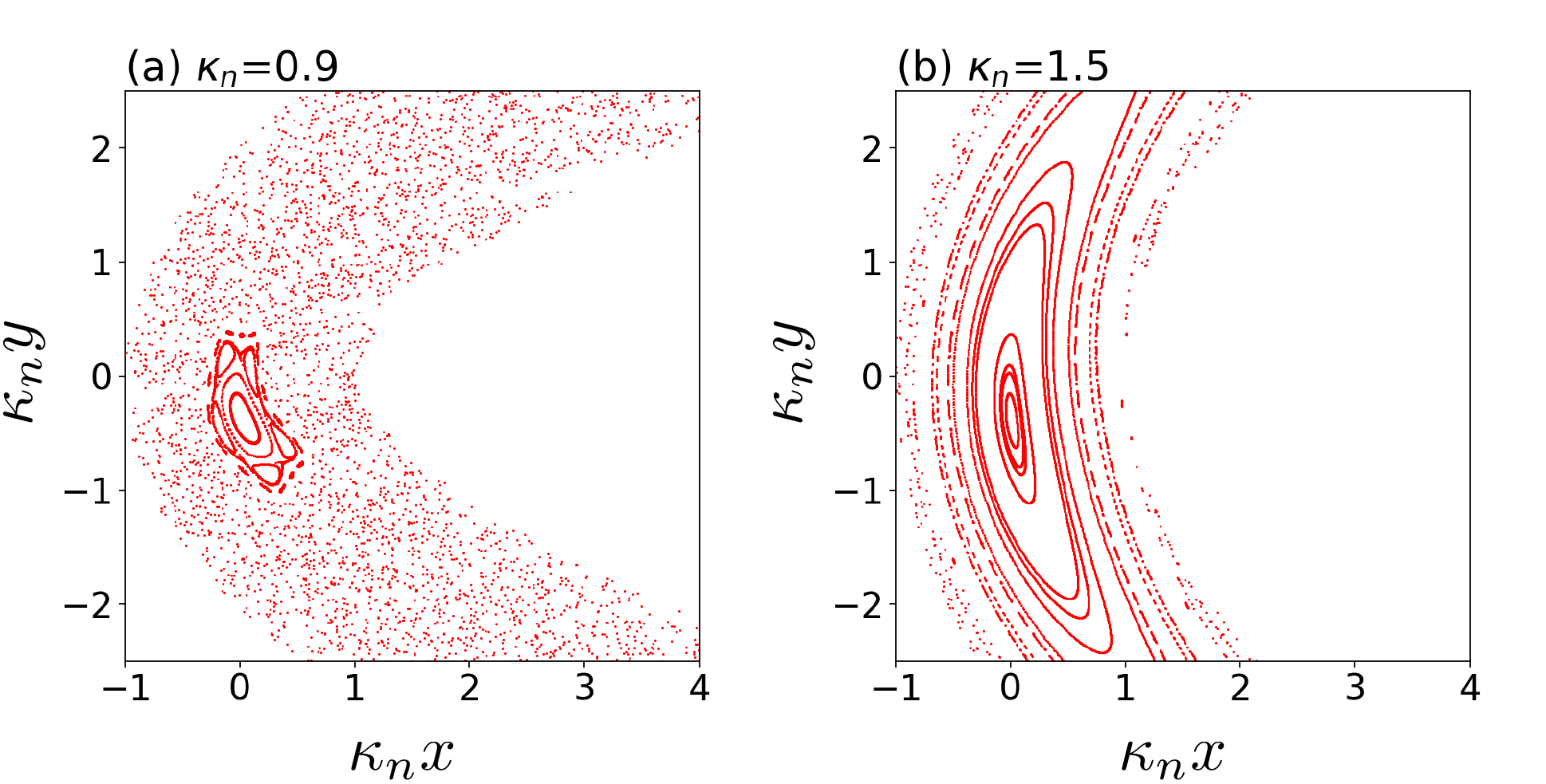}
\caption{(Color online)
\label{fig:poincare}
Poincar\'{e} surface of section plots at $v_z=0$
for (a) $\kappa_n=0.9$ and (b) $\kappa_n=1.5$
in the Box 3 model.
}
\end{figure}

We study the regime of the electron motion in Box 3,
by means of Poincar\'{e} surface of section plots.
The equation of electron motion in the shear field 
$\boldsymbol{B}(z) = B_0(z/L)\boldsymbol{e_{x}} + B_s \boldsymbol{e_{y}} + B_n \boldsymbol{e_{z}}$
can be normalized to
$\ddot{x} = -\kappa_n \dot{y} + \kappa_s \dot{z}$,~
$\ddot{y} = -z\dot{z} + \kappa_n \dot{x}$, and
$\ddot{z} = z\dot{y} - \kappa_s \dot{x}$.\citep{BZ91}
The shear curvature parameter is fixed to
$\kappa_s/\kappa_n = B_s/B_n = -1.25$.
This means $\kappa_{\rm tot} \approx 2.0 \kappa_n$.
The velocity is normalized to $\dot{x}^2+\dot{y}^2+\dot{z}^2=1$.
We set the initial position to
$(\kappa_n x, \kappa_n y, \kappa_n z)=( \dot{y}, -\dot{x}, 0)$
to adjust the canonical momentum and the constant of motion. 
The surface of section plots are taken
at negative-to-positive crossings at $\dot{z}=0$.
Figure \ref{fig:poincare} shows Poincar\'{e} maps for two cases. 
The corresponding parameters are also indicated by
the gray dashed lines in Figure \ref{fig:VDF}(h). 
In Figure \ref{fig:poincare}(a),
one can see the sea of chaos outside the regular core structure
for $\kappa_n = 0.9$.
The regular core region appears for $\kappa_n \gtrsim 0.75$, and
it corresponds to the adiabatic electrons.
The adiabatic electrons have small parallel velocities.
They gyrate fast but slowly bounce in the parallel direction
near the field reversal $z = 0$.
The electrons in the chaos region have high parallel velocities.
They move so fast in the parallel direction
that the field structure substantially changes in one gyroperiod, and
therefore the adiabaticity no longer holds. 
The map (Fig.~\ref{fig:poincare}(a)) confirms that
most of electrons are in the nonadiabatic regime
for $\kappa_{\rm tot} = 1.8$, the typical value in Box 3.
In contrast, Figure \ref{fig:poincare}(b) for $\kappa_{\rm n} = 1.5$
is occupied by the regular structure.
The chaos disappears and the electron motion becomes adiabatic
for $\kappa_{\rm n} \gtrsim 1.5$ or $\kappa_{\rm tot} \gtrsim 3.0$,
as mentioned in \citet{BZ91a}.

\end{document}